\documentclass[preprint,aps,amsmath,showpacs,showkeys,nofootinbib,superscriptaddress]{revtex4} 
\usepackage{multirow} 

\begin{document} 

\title{$\Omega NN$ and $\Omega\Omega N$ states}
\author{H.~Garcilazo} 
\email{humberto@esfm.ipn.mx} 
\affiliation{Escuela Superior de F\' \i sica y Matem\'aticas, \\ 
Instituto Polit\'ecnico Nacional, Edificio 9, 
07738 M\'exico D.F., Mexico} 

\author{A.~Valcarce} 
\email{valcarce@usal.es} 
\affiliation{Departamento de F\'\i sica Fundamental e IUFFyM,\\ 
Universidad de Salamanca, E-37008 Salamanca, Spain}

\date{\today} 

\begin{abstract} 
The lattice QCD analysis of the HAL QCD Collaboration has recently
derived $\Omega N$ and $\Omega\Omega$ interacting potentials with nearly physical
quark masses ($m_\pi \simeq $ 146 MeV and $m_K \simeq $ 525 MeV). 
They found an attractive interaction in the $\Omega N$ $^5S_2$ 
channel which supports a bound state with a central binding energy of 1.54 MeV.
The $\Omega \Omega$ $^1S_0$ channel shows an overall attraction with 
a bound state with a central binding energy of 1.6 MeV. 
In this paper we looked closely at the
$\Omega NN$ and $\Omega\Omega N$ three-body systems making use of the 
latest HAL QCD Collaboration $\Omega N$ and $\Omega\Omega$ interactions. 
Our results show that the $\Omega d$ system in
the state with maximal spin $(I)J^P=(0)5/2^+$ is bound
with a binding energy of about 20 MeV. The $(I)J^P=(1)3/2^+$
$\Omega nn$ state presents a resonance decaying to 
$\Lambda \Xi n$ and $\Sigma \Xi n$, 
with a separation energy of $\sim$ 1 MeV. The $(I)J^P=(1/2)1/2^+$
$\Omega \Omega N$ state also exhibits a resonance 
decaying to $\Lambda \Xi \Omega$ and $\Sigma \Xi \Omega$, 
with a separation energy of $\sim$ 4.6 MeV. 
We have calculated the contribution of the Coulomb potential 
to differentiate among the different charged states.
\end{abstract}



\maketitle 

\section{Introduction}
Few-body systems containing nucleons may enhance the binding 
of two-body resonances or bound states. We have simple examples
in nature. Whereas there is no evidence for strangeness $-1$ dibaryon 
states, the hypertriton $^3_\Lambda$H, $(I)J^P=(0)1/2^+$, is bound with 
a separation energy of $130 \pm 50$ keV, and the $^4_\Lambda$H, $(I)J^P=(0)0^+$, is bound
with a separation energy of $2.04 \pm 0.04$ MeV. Similarly, in the nonstrange sector
the binding per nucleon, $B/A$, increases from $1 : 3 : 7$ for increasing number of 
nucleons, $A= 2, 3 , 4$. Thus, the study of few-body systems could help in the
search for signals of two-body bound states or resonances in the strange sector.

The possible existence of dibaryons is a challenge since a long time 
ago~\cite{Dys64,Jaf77,Mul80}. Their occurrence would clearly help us to better 
understand Quantum Chromodinamycs in the nonperturbative regime.
Unfortunately, despite innumerable experimental searches, the deuteron
has been the only known dibaryon state up to very recently. 
The increasing quality of experimental data together with
exclusive measurements in hadronic reactions~\cite{Adl11},
direct measurements in relativistic heavy-ion collisions~\cite{Cho17},
or the development of new methods for nuclear emulsion experiments~\cite{Nak15},
has renewed the interest in dibaryons. Thus, the so--called WASA dibaryon,
a resonance 90 MeV below the nominal $\Delta\Delta$ threshold with a width of
70 MeV, is now firmly established although its nature is still under debate~\cite{Cle17}.
Moreover, the so--called KISO event represents the first clear evidence of a deeply bound 
state of $\Xi^{-} - ^{14}$N~\cite{Nak15}, which combined
with other indications of emulsion data suggest 
an average attractive $\Xi N$ interaction~\cite{Rij16}.
In this latter case, the increase in the binding energy per baryon 
within few-body systems with additional nucleons has already been noticed. 
If the $(I)J^P=(1)1^+$ $\Xi N$ state would be bound by 1.56 MeV~\cite{Nak15,Rij16}, 
the $\Xi d$ $(I)J^P=(1/2)3/2^+$ state becomes bound by 17.2 MeV~\cite{Gac16,Fil17}. 

In addition to improvements in the experimental 
data and techniques, the theoretical efforts of 
the lattice HAL QCD Collaboration~\cite{Aok12}
have reached the point of deriving baryon-baryon interactions near the physical
pion mass. Their most recent results hint to the
existence of shallow bound states in the $\Omega N$ and $\Omega\Omega$ 
systems~\cite{Gon18,Iri18}. Two-body systems containing
an $\Omega$ baryon seem to be specially suited to lodge a two-body
bound state or resonance. Thus, the $\Omega N$ interaction is expected to 
lack a repulsive core since the quark flavors of the nucleon are different
from those of the $\Omega$, so that the Pauli exclusion
principle can not act~\cite{Eti14,Gol87}. Moreover, the color magnetic
term of the one-gluon exchange interaction is attractive in some 
particular channels, one example being the $^5S_2$, what led to the 
prediction of an $\Omega N$ dibaryon in constituent quark models~\cite{Gol87,Hua15}.
The recent lattice QCD results by the HAL QCD Collaboration hint toward
the existence of such $\Omega N$ bound state~\cite{Iri18}.
Another two-body system containing $\Omega$ baryons, $\Omega\Omega$, is 
also interesting because it would be the only possible
strong interaction stable state made of two decuplet baryons.
The adequacy of such two-body system for having an overall attractive interaction has become apparent
on phenomenological quark models predicting an $\Omega\Omega$ $^1S_0$ bound 
state~\cite{Zha00,Dai06}. The recent near to physical pion mass results
of the lattice HAL QCD Collaboration corroborate such finding~\cite{Gon18}.
It is worth to notice that the theoretical analysis~\cite{Mor16} of the 
first measurement of the proton-$\Omega$ correlation function in heavy-ion collisions
by the STAR experiment~\cite{Ada18} at the Relativistic Heavy-Ion Collider (RHIC)
favors the proton-$\Omega$ bound state hypothesis~\footnote{Note, however,
that the first measurement of the proton-$\Omega$ correlation function drives
to a proton-$\Omega$ bound state with a binding energy of $\sim$ 27 MeV~\cite{Ada18}, 
far from the present lattice QCD results of about 1.54 MeV~\cite{Iri18} that will be 
used in this work.}. 

Preliminary results by the HAL QCD 
Collaboration~\cite{Doi18} had already shown the attractive character 
of the $^5S_2$ $\Omega N$ interaction. These 
exploratory results were parametrized in Ref.~\cite{Sek18} by means of an equivalent
local potential reproducing the $\Omega N$ $^5S_2$ scattering amplitude
of Ref.~\cite{Doi18}. Making use of the $\Omega N$ equivalent local potential 
of Ref.~\cite{Sek18} together with the Malfliet-Tjon $^3S_1$ $NN$ potential of Ref.~\cite{Fri90}, 
we studied the $\Omega d$ system in the maximal spin channel,
$(I)J^P=(0)5/2^+$. We concluded that the system is bound, 
with a binding energy of about 17 MeV~\cite{Gar18}. The 
latest results of the HAL QCD Collaboration nearly physical quark
masses ($m_\pi \simeq $ 146 MeV and $m_K \simeq $ 525 MeV) have been recently
made public~\cite{Iri18}, validating the attractive nature of the $\Omega N$
$^5S_2$ state and providing their own parametrization of the lattice QCD
interaction. Moreover, the HAL QCD Collaboration has also recently 
published the $\Omega\Omega$ $^1S_0$ interaction
nearly physical pion mass, $m_\pi =$ 146 MeV, 
showing its overall attractive character~\cite{Gon18}. 
Thus, we have all the necessary ingredients to explore possible implications of 
the attractive nature of the $\Omega N$ and $\Omega \Omega$ interactions on
few-body systems on the basis of first-principle lattice QCD-based interactions.

Our purpose in this work is to study the
three-body systems $\Omega NN$ and $\Omega \Omega N$ looking for
deeply bound states or resonances which may be sought experimentally. 
The paper is organized as follows. In Sec.~\ref{secII} we summarize the two-body
interactions. In Sec.~\ref{secIII} we outline the solution of the
three-body bound-state Faddeev equations for the case of two identical particles.
In Sec.~\ref{secIV} we present and discuss our results. The most important
conclusions of our work are summarized in Sec.~\ref{secV}.

\section{Two-body interactions}
\label{secII}
In the case of the $NN$ subsystem the two-body potential
in configuration space is given by~\cite{Fri90},
\begin{equation}
V_{NN}(r)=\sum_{n=1}^2C_n \frac{e^{-\mu_n r}}{r} \, ,
\label{eq77}
\end{equation}
with the parameters specified in Table~\ref{tab0}. The predicted binding energy of the
deuteron is $B=$ 2.2307 MeV.
\begin{table}[t]
\caption{Low-energy data and parameters of the local central Yukawa-type potential 
given by Eq.~(\ref{eq77}) for the $^1S_0$ and $^3S_1$ $NN$ interactions~\cite{Fri90}.} 
\begin{ruledtabular} 
\begin{tabular}{ccccccc} 
 $(I,J)$ & $a_0$ (fm) & $r_{\rm eff}$ (fm) & $C_1$ (MeV fm) & 
$\mu_1$ (fm$^{-1}$) 
& $C_2$ (MeV fm) & $\mu_2$ (fm$^{-1}$)  \\
\hline
 $(1,0)$  & $-23.56$ & $2.88$ & $-513.968$  & $1.55$  & $1438.72$ & $3.11$ \\
 $(0,1)$  & $5.51$ & $1.89$ & $-626.885$  & $1.55$  & $1438.72$ & $3.11$ \\
\end{tabular}
\end{ruledtabular}
\label{tab0} 
\end{table}

For the $\Omega N$ system the HAL QCD Collaboration has recently published
the $S$-wave and spin 2 interaction with nearly physical quark 
masses~\cite{Iri18}. The lattice results are fitted by an analytic
function composed of an attractive Gaussian core plus
a long range $($Yukawa$)^2$ attraction with a form factor~\cite{Eti14},
\begin{equation}
V_{\Omega N} (r) = b_1 e^{-b_2 \, r^2} \, + \, b_3 \left( 1 - e^{-b_4 \, r^2}\right)
\left(\frac{e^{-m_\pi \, r}}{r}\right) ^2 \, .
\label{eqq1}
\end{equation}
The $($Yukawa$)^2$ form at long distance is motivated by the picture of two-pion
exchange between $N$ and $\Omega$ with an OZI violating vertex~\cite{Sek18}. The pion
mass in Eq.~(\ref{eqq1}) is taken from the simulation, $m_\pi =$ 146 MeV~\cite{Iri18}.
The lattice results are fitted reasonably well,
$\chi^2/$d.o.f $\simeq$ 1, with four different set of parameters given in 
Table~\ref{tab1}. The above interactions drive to the following central values
of the low-energy data: a scattering length $a_0^{\Omega N} =$ 5.30 fm, an effective range
$r_{\rm eff}^{\Omega N} =$ 1.26 fm, and a binding energy $B_{\Omega N}$ = 1.54 MeV
\begin{table}[t]
\caption{Fitting parameters in Eq.~(\ref{eqq1}) for different models, $P_i$, of the
$^5S_2$ $\Omega N$ interaction~\cite{Iri18}.}\label{tab1}
\begin{ruledtabular}
\begin{tabular}{l|rrrr}
 & $P_1$ & $P_2$ & $P_3$ & $P_4$ \\
\hline
$b_1$ (MeV)            &  $-$306.5  &  $-$313.0  &  $-$316.7  & $-$296  \\
$b_2$ (fm$^{-2}$)      &      73.9  &      81.7  &      81.9  &     64  \\
$b_3$ (MeV fm$^{2}$)   &  $-$266    &  $-$252    &  $-$237    & $-$272  \\
$b_4$ (fm$^{-2}$)      &       0.78 &       0.85 &       0.91 &      0.76 \\
\end{tabular}
\end{ruledtabular}
\end{table}

Finally, the $\Omega\Omega$ interaction in the $^1S_0$ channel has also been recently
studied by the HAL QCD Collaboration with a large volume and nearly the physical pion
mass~\cite{Gon18}. The results indicate that the $\Omega \Omega$ interaction has an
overall attraction. In particular, the potential derived has been fitted by means of an
analytic function composed of three Gaussians,
\begin{equation}
V_{\Omega\Omega}(r)=\sum_{j=1}^3 c_j e^{-(r/d_j)^2} \, ,
\label{eqq7}
\end{equation}
where $(c_1,c_2,c_3)=(914,305,-112)$ in MeV and 
$(d_1,d_2,d_3)=(0.143,0.305,0.949)$ in fm with
$\chi^2/$d.o.f $\simeq$ 1.3. A functional form similar to the
one used for the $\Omega N$ interaction, i.e., two Gaussians 
plus a long range $($Yukawa function$)^2$ provides an equally good
fit and does not affect the results. The low-energy data obtained with
this interaction are the following: a scattering length 
$a_0^{\Omega\Omega} =$ 4.6 fm, an effective range
$r_{\rm eff}^{\Omega\Omega} =$ 1.27 fm, and a binding energy $B_{\Omega\Omega}$ = 1.6 MeV.

\section{The three-body bound-state Faddeev equations}
\label{secIII}

The $\Omega NN$ and $\Omega\Omega N$ three-body problems are studied 
using the method of Ref.~\cite{Gar16}, expanding the two-body amplitudes 
in terms of Legendre polynomials. 
We restrict ourselves to configurations where all three 
particles are in $S$-wave states so that the Faddeev equations 
for the bound-state problem in the case of three baryons with total
isospin $I$ and spin $J$ are,
\begin{eqnarray}
T_{i;IJ}^{i_ij_i}(p_iq_i) = &&\sum_{j\ne i}\sum_{i_jj_j}
h_{ij;IJ}^{i_ij_i;i_jj_j}\frac{1}{2}\int_0^\infty q_j^2dq_j
\int_{-1}^1d{\rm cos}\theta\, 
t_{i;i_ij_i}(p_i,p_i^\prime;E-q_i^2/2\nu_i) 
\nonumber \\ &&
\times\frac{1}{E-p_j^2/2\mu_j-q_j^2/2\nu_j}\;
T_{j;IJ}^{i_jj_j}(p_jq_j) \, , 
\label{eq1} 
\end{eqnarray}
where $t_{i;i_ij_i}$ stands for the two-body amplitudes
with isospin $i_i$ and spin $j_i$. $p_i$
is the momentum of the pair $jk$ (with $ijk$ an even permutation of
$123$) and $q_i$ the momentum of particle $i$ with respect to the pair
$jk$. $\mu_i$ and $\nu_i$ are the corresponding reduced masses,
\begin{eqnarray}
\mu_i &=& \frac{m_jm_k}{m_j+m_k} \, , \nonumber\\
\nu_i &=& \frac{m_i(m_j+m_k)}{m_i+m_j+m_k} \, ,
\label{eq3}
\end{eqnarray}
and the momenta $p_i^\prime$ and $p_j$ in Eq.~(\ref{eq1}) are given by,
\begin{eqnarray}
p_i^\prime &=& \sqrt{q_j^2+\frac{\mu_i^2}{m_k^2}q_i^2+2\frac{\mu_i}{m_k}
q_iq_j{\rm cos}\theta} \, , \nonumber \\
p_j &=& \sqrt{q_i^2+\frac{\mu_j^2}{m_k^2}q_j^2+2\frac{\mu_j}{m_k}
q_iq_j{\rm cos}\theta} \, .
\label{eq5}
\end{eqnarray}
$h_{ij;IJ}^{i_ij_i;i_jj_j}$ are the spin--isospin coefficients,
\begin{eqnarray}
h_{ij;IJ}^{i_ij_i;i_jj_j}= &&
(-)^{i_j+\tau_j-I}\sqrt{(2i_i+1)(2i_j+1)}
W(\tau_j\tau_kI\tau_i;i_ii_j)
\nonumber \\ && \times
(-)^{j_j+\sigma_j-J}\sqrt{(2j_i+1)(2j_j+1)}
W(\sigma_j\sigma_kJ\sigma_i;j_ij_j) \, , 
\label{eq6}
\end{eqnarray}
where $W$ is the Racah coefficient and $\tau_i$, $i_i$, and $I$ 
($\sigma_i$, $j_i$, and $J$) are the isospins (spins) of particle $i$,
of the pair $jk$, and of the three--body system.

Since the variable $p_i$ in Eq.~(\ref{eq1}) runs from 0 to $\infty$,
it is convenient to make the transformation
\begin{equation}
x_i=\frac{p_i-b}{p_i+b} \, ,
\label{eq7}
\end{equation}
where the new variable $x_i$ runs from $-1$ to $1$ and $b$ is a scale
parameter that has no effect on the solution. With this transformation
Eq.~(\ref{eq1}) takes the form,
\begin{eqnarray}
T_{i;IJ}^{i_ij_i}(x_iq_i) = &&\sum_{j\ne i}\sum_{i_jj_j}
h_{ij;IJ}^{i_ij_i;i_jj_j}\frac{1}{2}\int_0^\infty q_j^2dq_j
 \int_{-1}^1d{\rm cos}\theta\; 
t_{i;i_ij_i}(x_i,x_i^\prime;E-q_i^2/2\nu_i) 
\nonumber \\ &&
\times\frac{1}{E-p_j^2/2\mu_j-q_j^2/2\nu_j}\;
T_{j;IJ}^{i_jj_j}(x_jq_j) \, . 
\label{eq8} 
\end{eqnarray}
Since in the amplitude $t_{i;i_ij_i}(x_i,x_i^\prime;e)$ the variables
$x_i$ and $x_i^\prime$ run from $-1$ to $1$, one can expand this amplitude
in terms of Legendre polynomials as,
\begin{equation}
t_{i;i_ij_i}(x_i,x_i^\prime;e)=\sum_{nr}P_n(x_i)\tau_{i;i_ij_i}^{nr}(e)P_r(x'_i) \, ,
\label{eq9}
\end{equation}
where the expansion coefficients are given by,
\begin{equation}
\tau_{i;i_ij_i}^{nr}(e)= \frac{2n+1}{2}\frac{2r+1}{2}\int_{-1}^1dx_i
\int_{-1}^1 dx_i^\prime\; P_n(x_i) 
t_{i;i_ij_i}(x_i,x_i^\prime;e)P_r(x_i^\prime) \, .
\label{eq10} 
\end{equation}
Applying expansion~(\ref{eq9}) in Eq.~(\ref{eq8}) one gets,
\begin{equation}
T_{i;IJ}^{i_ij_i}(x_iq_i) = \sum_n P_n(x_i) T_{i;IJ}^{ni_ij_i}(q_i) \, ,
\label{eq11}
\end{equation}
where $T_{i;IJ}^{ni_ij_i}(q_i)$ satisfies the one-dimensional integral equation,
\begin{equation}
T_{i;IJ}^{ni_ij_i}(q_i) = \sum_{j\ne i}\sum_{mi_jj_j}
\int_0^\infty dq_j A_{ij;IJ}^{ni_ij_i;mi_jj_j}(q_i,q_j;E)\;
T_{j;IJ}^{mi_jj_j}(q_j) \, , 
\label{eq12}
\end{equation}
with
\begin{eqnarray}
A_{ij;IJ}^{ni_ij_i;mi_jj_j}(q_i,q_j;E)= &&
h_{ij;IJ}^{i_ij_i;i_jj_j}\sum_r\tau_{i;i_ij_i}^{nr}(E-q_i^2/2\nu_i)
\frac{q_j^2}{2}
\nonumber \\ &&
\times\int_{-1}^1 d{\rm cos}\theta\;\frac{P_r(x_i^\prime)P_m(x_j)} 
{E-p_j^2/2\mu_j-q_j^2/2\nu_j} \, .
\label{eq13} 
\end{eqnarray}

The three amplitudes $T_{1;IJ}^{ri_1j_1}(q_1)$, $T_{2;IJ}^{mi_2j_2}(q_2)$,
and $T_{3;IJ}^{ni_3j_3}(q_3)$ in Eq.~(\ref{eq12}) are coupled together.
The number of coupled equations can be reduced, however, when two of
the particles are identical, which is currently the case, $\Omega NN$ and
$\Omega\Omega N$. The procedure for the case of two identical fermions 
has been described before~\cite{Afn74,Gar90}
and will not be repeated here. With the assumption that 
particles 2 and 3 are identical and
particle 1 is the different one, only the amplitudes 
$T_{1;IJ}^{ri_1j_1}(q_1)$ and $T_{2;IJ}^{mi_2j_2}(q_2)$ are independent
from each other and they satisfy the coupled integral equations,
\begin{equation}
T_{1;IJ}^{ri_1j_1}(q_1)  =  2\sum_{mi_2j_2}
\int_0^\infty dq_3 A_{13;IJ}^{ri_1j_1;mi_2j_2}(q_1,q_3;E)\;
T_{2;IJ}^{mi_2j_2}(q_3) \, ,
\label{eq14} 
\end{equation}
\begin{eqnarray}
T_{2;IJ}^{ni_2j_2}(q_2) = && \sum_{mi_3j_3}g
\int_0^\infty dq_3 A_{23;IJ}^{ni_2j_2;mi_3j_3}(q_2,q_3;E)\;
T_{2;IJ}^{mi_3j_3}(q_3) 
\nonumber \\ && +
\sum_{ri_1j_1}
\int_0^\infty dq_1 A_{31;IJ}^{ni_2j_2;ri_1j_1}(q_2,q_1;E)\;
T_{1;IJ}^{ri_1j_1}(q_1) \, , 
\label{eq15} 
\end{eqnarray}
with the identical--particle factor
\begin{equation}
g=(-)^{1+\sigma_1+\sigma3-j_2+\tau_1+\tau_3-i_2} \, ,
\label{eq16}
\end{equation}
where $\sigma_1$ ($\tau_1$) 
stand for the spin (isospin) of the different particle and
$\sigma_3$ ($\tau_3$) for those of the identical ones.

Substitution of Eq.~(\ref{eq14}) into Eq.~(\ref{eq15}) yields an
equation with only the amplitude $T_2$,
\begin{equation}
T_{2;IJ}^{ni_2j_2}(q_2) = \sum_{mi_3j_3}
\int_0^\infty dq_3 K_{IJ}^{ni_2j_2;mi_3j_3}(q_2,q_3;E)\;
T_{2;IJ}^{mi_3j_3}(q_3) \, , 
\label{eq17}
\end{equation}
where
\begin{eqnarray}
K_{IJ}^{ni_2j_2;mi_3j_3}(q_2,q_3;E)= && g
A_{23;IJ}^{ni_2j_2;mi_3j_3}(q_2,q_3;E)
\nonumber \\ && +
2\sum_{ri_1j_1}
\int_0^\infty dq_1 A_{31;IJ}^{ni_2j_2;ri_1j_1}(q_2,q_1;E)
A_{13;IJ}^{ri_1j_1;mi_3j_3}(q_1,q_3;E) \, .
\label{eq18} 
\end{eqnarray}

The off-shell two-body $t-$matrices are obtained by solving
the Lippmann-Schwinger equation,
\begin{equation}
t_i(p_i,p_i^\prime;e)= V_i(p_i,p_i^\prime)+ \int_0^\infty 
{p_i^{\prime\prime}}^2dp_i^{\prime\prime}
V_i(p_i,p_i^{\prime\prime})
\frac{1}{e-{p_i^{\prime\prime}}^2/2\eta_i+i\epsilon} 
t_i(p_i^{\prime\prime},p_i^\prime;e) \, .
\label{eq66}
\end{equation}
with the two-body interactions $V_i$, $i=NN,\Omega N, \Omega\Omega$, described 
in section~\ref{secII}.

Finally, in order to separate the binding energies of the different charged 
states we included the Coulomb interaction as,
\begin{equation}
V_C(r)=\pm\alpha \, \frac{e^{-r/r_0}}{r} \, ,
\label{eqcou}
\end{equation}
where $\alpha$ is the fine structure constant and $r_0$ a screening radius
taken to be $r_0=50$ fm.

\section{Results}
\label{secIV}
Let us first of all present and discuss our results for the corresponding 
$\Omega NN$ states\footnote{We have not considered the channel coupling
to lower inelastic channels and, therefore, we have not estimated the width
of the states having a lower decay channel different from those made
of $N$'s and $\Omega$'s.}. We have first studied the $\Omega d$ state. We show in 
Table~\ref{tab3} the binding energies of the state with maximal spin,
$(I)J^P=(0)5/2^+$, for the different models of the $\Omega N$ interaction 
reported in Ref.~\cite{Iri18} and summarized in Table~\ref{tab1}.
For completeness we also include the binding energy of the
$^5S_2$ $\Omega N$ state. 
As indicated in the caption, the numbers between parenthesis correspond to
the results using the N and $\Omega$ masses derived by the HAL QCD
Collaboration~\cite{Gon18,Iri18}, that are somewhat larger than the experimental masses.
If the masses increase, the repulsive kinetic energy contribution decreases,
resulting in slightly larger binding energies. 
As it can be seen, the binding energy of the $\Omega N$ system is larger than in the
preliminary results of the $\Omega N$ interaction presented in Ref.~\cite{Doi18}
and parametrized in Ref.~\cite{Sek18}, where the $\Omega N$ binding energy was 
0.3 MeV~\cite{Sek18,Gar18} for the $N$ and $\Omega$ physical masses.
As a consequence, the binding of the $\Omega d$ state with maximal spin  
$(I)J^P=(0)5/2^+$ also increases from the 16.34 MeV measured with respect to the
$\Omega np$ threshold~\cite{Gar18} to about 20 MeV. 
Let us emphasize that the $\Omega d$ in the maximal spin channel $(I)J^P=(0)5/2^+$
cannot couple to the lower channels $\Lambda\Xi N$ and $\Sigma\Xi N$
with the $\Lambda\Xi$ and $\Sigma\Xi$ subsystems in $S$ waves, so that the
width of a $\Omega d$ bound state is expected to be small.
\begin{table}[t]
\caption{Binding energy of the $^5S_2$ $\Omega N$ state, $B_{\Omega N}$, and
the $(I)J^P=(0)5/2^+$ $\Omega d$ state, $B_{\Omega d}$, for the different models of the
$\Omega N$ interaction given in Table~\ref{tab1}~\cite{Iri18}. 
The results have been obtained with
the experimental masses of the $N$ and $\Omega$, 938.9 MeV/c$^2$ and 1672.45 MeV/c$^2$
respectively. We have indicated between parenthesis the results corresponding
to the N and $\Omega$ masses derived by the HAL QCD
Collaboration, 954.7 MeV/c$^2$ and 1711.5 MeV/c$^2$ respectively~\cite{Gon18,Iri18}.
All energies are in MeV.}\label{tab3}
\begin{ruledtabular}
\begin{tabular}{c|cccc}
  & $P_1$ & $P_2$ & $P_3$ & $P_4$ \\
\hline
$B_{\Omega N}$  &  1.29 (1.52)   &  1.38 (1.61)   &  1.29 (1.44)   &  1.37 (1.60)  \\
$B_{\Omega d}$  &  19.6 (20.6)   &  20.0 (21.1)   &  19.6 (20.5)   &  19.9 (20.9) \\
\end{tabular}
\end{ruledtabular}
\end{table}
\begin{table}[b]
\caption{Binding energy, $B_{\Omega nn}$, and separation energy, $S_{\Omega nn}$, of the
the $(I)J^P=(1)3/2^+$ state for the different models of the
$\Omega N$ interaction given in Table~\ref{tab1}~\cite{Iri18}. All energies are in MeV.
The numbers between parenthesis have the same meaning as in Table~\ref{tab3}.}\label{tab4}
\begin{ruledtabular}
\begin{tabular}{c|cccc}
  & $P_1$ & $P_2$ & $P_3$ & $P_4$ \\
\hline
$B_{\Omega nn}$ &  2.25 (2.60)  &  2.35 (2.72)  &  2.14 (2.50)  &  2.34 (2.71)  \\
$S_{\Omega nn}$ &  0.96 (1.08)  &  0.97 (1.11)  &  0.85 (1.06)  &  0.97 (1.11) \\
\end{tabular}
\end{ruledtabular}
\end{table}

Regarding the $\Omega nn$ ($\Omega pp$) system, with the $\Omega N$ interaction derived in 
Ref.~\cite{Iri18} one can construct a $\Omega nn$ ($\Omega pp$) state with quantum numbers $(I)J^P=(1)3/2^+$.
The results obtained for this state with the different models of the $\Omega N$ interaction
given in Table~\ref{tab1} are shown in Table~\ref{tab4}.
Such $\Omega nn$ state could decay to  $\Lambda \Xi n$ and $\Sigma \Xi n$, 
thus it would appear as a resonance.
\begin{table}[t]
\caption{Binding energy, $B_{\Omega \Omega N}$, and separation energy, $S_{\Omega \Omega N}$, of the
the $(I)J^P=(1/2)1/2^+$ $\Omega \Omega N$ state for the different models of the
$\Omega N$ interaction given in Table~\ref{tab1}~\cite{Iri18} and the $\Omega\Omega$ interaction
of Ref.~\cite{Gon18}. All energies are in MeV.
The numbers between parenthesis have the same meaning as in Table~\ref{tab3}.}\label{tab5}
\begin{ruledtabular}
\begin{tabular}{c|cccc}
  & $P_1$ & $P_2$ & $P_3$ & $P_4$ \\
\hline
$B_{\Omega \Omega N}$ &  6.0 (6.7)  &  6.2 (6.9)  &  5.9 (6.6)  &  6.1 (6.8)  \\
$S_{\Omega \Omega N}$ &  4.6 (5.1)  &  4.8 (5.3)  &  4.5 (5.0)  &  4.7 (5.2) \\
\end{tabular}
\end{ruledtabular}
\end{table}

We present now the results obtained for the $\Omega\Omega N$ system using the $\Omega N$ potential 
derived in Ref.~\cite{Iri18} and the $\Omega\Omega$ interaction reported in Ref.~\cite{Gon18},
both for the same pion mass. The two partial waves analyzed by the HAL QCD Collaboration
generate an $\Omega\Omega N$ three-body system with quantum numbers $(I)J^P=(1/2)1/2^+$.
This state is bound for all parametrizations of the $\Omega N$ interaction, with the binding
energy, $B_{\Omega\Omega N}$, and separation energy, $S_{\Omega\Omega N}$, given in Table~\ref{tab5}.
Note that the binding energy of the $^1S_0$ $\Omega \Omega$ state is 1.4 MeV for the physical mass
of the $\Omega$ baryon, and 1.6 MeV for the HAL QCD mass of the $\Omega$ baryon.
This state would also appear as a resonance decaying to $\Lambda \Xi \Omega$ and $\Sigma \Xi \Omega$.

We have calculated the contribution of the Coulomb potential exactly as dictated by Eq.~(\ref{eqcou}),
in order to differentiate among the different
charged states of the $\Omega N$, $\Omega NN$ and $\Omega\Omega N$ systems. The Coulomb 
increases the binding for systems containing a proton compared to those with
a neutron, due to the attractive $\Omega^- p$ interaction. 
The Coulomb effects on the binding energy of the 
$^5S_2$ $\Omega N$ state were estimated in Ref.~\cite{Iri18}, 
obtaining a difference of 0.92 MeV between the $\Omega p$ and $\Omega n$ states.
We have checked this result for the different models, $P_i$, of the $\Omega N$ interaction
by means of the potential given in Eq.~(\ref{eqcou}). We have obtained
corrections of: 0.90, 0.92, 0.90 and 0.91 MeV respectively, in agreement
with the lattice QCD results.
In the $(I)J^P=(1)3/2^+$ $\Omega NN$ system the Coulomb potential induces an extra binding
of 0.7 MeV for the $\Omega pp$ state as compared to the $\Omega nn$ one. In the
$(I)J^P=(0)5/2^+$ $\Omega d$ state the Coulomb interaction generates an additional 
binding of about 1.3 MeV. For the $\Omega\Omega$ $^1S_0$ state the Coulomb interaction reduces
the binding energy from the 1.6 MeV reported in Ref.~\cite{Gon18} to 0.9 MeV. Finally, 
in the $\Omega\Omega N$ system the Coulomb potential
induces an extra binding of 0.3 MeV for the $\Omega\Omega p$ state whereas it
penalizes the $\Omega\Omega n$ state by 1 MeV, generating a splitting of 1.3 MeV between
these two states. The final results are summarized in Table~\ref{tab6}.
\begin{table}[t]
\caption{Binding energy of the different $\Omega N$, $\Omega N N$ and $\Omega \Omega N$ charged states
including the Coulomb potential given by Eq.~(\ref{eqcou}). All energies are in MeV.
The numbers between parenthesis have the same meaning as in Table~\ref{tab3}.}\label{tab6}
\begin{ruledtabular}
\begin{tabular}{cc|cccc}
$(I)J^P$  & System & $P_1$ & $P_2$ & $P_3$ & $P_4$ \\
\hline

\multirow{2}{*}{$(1/2)2^+$}   & $\Omega n$         & 1.29 (1.52)  &  1.38 (1.61)  &  1.29 (1.44) &  1.37 (1.60)  \\
                              & $\Omega p$         & 2.19 (2.42)  &  2.30 (2.53)  &  2.19 (2.34) &  2.28 (2.51)  \\
\multirow{2}{*}{$(1)3/2^+$}   & $\Omega nn$        & 2.25 (2.60)  &  2.35 (2.72)  &  2.14 (2.50) &  2.34 (2.71) \\
                              & $\Omega pp$        & 2.91 (3.30)  &  3.04 (3.43)  &  2.81 (3.20) &  3.02 (3.41) \\
\multirow{1}{*}{$(0)5/2^+$}   & $\Omega d$         & 20.9 (22.0)  &  21.3 (22.4)  &  20.7 (21.8) &  21.2 (22.3)\\	
\multirow{2}{*}{$(1/2)1/2^+$} & $\Omega \Omega n$  & 5.0 (5.6)    &  5.1 (5.8)    &  4.9 (5.5)   &  5.1 (5.8)  \\
                              & $\Omega \Omega p$  & 6.3 (7.0)    &  6.5 (7.2)    &  6.2 (6.9)   &  6.4 (7.2)  \\												
\end{tabular}
\end{ruledtabular}
\end{table}

Let us finally note that to draw definite conclusion about the
width of states with a lower decay channel, one should have done a coupled-channel
calculation. However, for this purpose one would need the transition potentials
to the inelastic channels that have still not been derived from lattice QCD
and thus the conclusions would be speculative. 
The width of a three-body resonance in a coupled two-channel system has been recently 
estimated in Ref.~\cite{Gar17}, presenting a plausible argument to explain the small 
width of a three-body resonance lying close to the upper channel in spite of being 
open the lower one. The analysis of the width of a two-body resonance in
a coupled-channel system has also been recently presented in Ref.~\cite{Gac18}. 
It has been demonstrated how the width
does not come only determined by the available phase space
for its decay to the detection channel, but it greatly depends
on the relative position of the mass of the resonance with
respect to the masses of the coupled-channels generating the
state. As seen in Fig. 1 of this reference,
the resonance may still be narrow being close to the upper threshold if there is a 
little overlap with the wave function of the lower inelastic channel.
Hence, in the region where the dynamics is dominated by the attraction 
in the upper channel, the resonance could still be narrow and the lower channel 
would be mainly a tool for the detection. 
This mechanism is somewhat related to the 'synchronization of resonances' proposed 
by D.~Bugg~\cite{Bug08}. Thus, our studies about the width of two- and three-body
resonances suggest the possibility of the experimental observation
of narrow resonances lying well above their lowest decay
threshold.

\section{Outlook}
\label{secV}
The lattice QCD analysis of the HAL QCD Collaboration has recently
derived $\Omega N$ and $\Omega\Omega$ interacting potentials with nearly physical
quark masses ($m_\pi \simeq $ 146 MeV and $m_K \simeq $ 525 MeV). 
They found an attractive potential in the $\Omega N$ $^5S_2$ 
channel which supports a bound state with a central binding energy of 1.54 MeV.
The $\Omega \Omega$ $^1S_0$ channel shows an overall attraction with 
a bound state with a central binding energy of 1.6 MeV. 
On the basis of our current understanding of the nonstrange sector, where the 
binding energy may increase with the number of baryons, in this paper we 
have examined carefully the
$\Omega NN$ and $\Omega\Omega N$ three-body systems making use of the 
latest HAL QCD Collaboration $\Omega N$ and $\Omega\Omega$ interactions.
We have looked for deeply bound states or resonances which may be sought experimentally.
Our results show that the $\Omega d$ system in
the state with maximal spin $(I,J^P)=(0,5/2^+)$ is bound
with a binding energy of about 20 MeV. 
The $\Omega d$ in the maximal spin channel $(I)J^P=(0)5/2^+$
cannot couple to the lower channels $\Lambda\Xi N$ and $\Sigma\Xi N$
with the $\Lambda\Xi$ and $\Sigma\Xi$ subsystems in $S$ waves, so that the
width of a $\Omega d$ bound state is expected to be small.
The $(I,J^P)=(1,3/2^+)$ $\Omega nn$ state presents a 
resonance decaying to $\Lambda \Xi n$ and $\Sigma \Xi n$, 
with a separation energy of $\sim$ 1 MeV. The 
$(I,J^P)=(1/2,1/2^+)$ $\Omega \Omega N$ state 
also exhibits a resonance decaying to $\Lambda \Xi \Omega$ and $\Sigma \Xi \Omega$, 
with a separation energy of $\sim$ 4.6 MeV. 
The Coulomb potential increases the binding for systems containing a proton 
compared to those with a neutron, due to the attractive $\Omega^- p$ interaction.
It also penalizes the binding of the $\Omega\Omega$ state. The overall effect is
of the order of 1 MeV, as noticed in the lattice QCD calculations of the two-body
systems.

The latest baryon-baryon interactions in the strange sector with nearly
physical pion mass based on lattice QCD hint toward the existence of 
bound states or sharp resonances. These states could be observed
by hadron beam experiments at J-PARC and FAIR, or by relativistic heavy-ion
collisions at RHIC and LHC. In Ref.~\cite{Mor16} it has been discussed
how the two-particle momentum correlation between the proton and the $\Omega$ baryon 
in high-energy heavy ion collisions may unveil the existence of these states.
The first measurement of the proton-$\Omega$ correlation function in heavy-ion collisions
by the STAR experiment~\cite{Ada18} at RHIC
favors the proton-$\Omega$ bound state hypothesis. We hope our theoretical studies 
could help to design experiments where 
these lattice QCD-based predictions could be tested. 

\begin{acknowledgments} 
This work has been partially funded by COFAA-IPN (M\'exico) and 
by Ministerio de Econom\'\i a, Industria y Competitividad 
and EU FEDER under Contract No. FPA2016-77177-C2-2-P.
\end{acknowledgments}

\end{document}